\documentclass[10pt,letterpaper]{article} 
\usepackage[top=0.85in,left=2.75in,footskip=0.75in]{geometry}
\usepackage{amsmath,amssymb}
\usepackage{changepage}

\usepackage{textcomp,marvosym}
\usepackage{subcaption}

\usepackage[T1]{fontenc}
\usepackage{lmodern}
\usepackage{cite}

\usepackage{nameref,hyperref}

\usepackage[right]{lineno}

\usepackage[nopatch=eqnum]{microtype}
\DisableLigatures[f]{encoding = *, family = * }

\usepackage[table]{xcolor}

\usepackage{array}
\usepackage{caption}
\newcolumntype{+}{!{\vrule width 2pt}}

\newlength\savedwidth

\raggedright
\usepackage[aboveskip=1pt,labelfont=bf,labelsep=period,justification=raggedright,singlelinecheck=off]{caption}

\newcommand{\fref}[1]{Fig~\ref{#1}}
\newcommand{\eref}[1]{Eq~(\ref{#1})}
\makeatletter
\renewcommand{\@biblabel}[1]{\quad#1.}
\makeatother

\usepackage{lastpage,fancyhdr,graphicx}
\usepackage{epstopdf}

\usepackage[english]{babel}
\usepackage{graphicx,helvet}
\usepackage{color}
\usepackage{url}
\usepackage{amssymb}
\usepackage[utf8]{inputenc}
\usepackage{hyperref}
\usepackage{bbm,bm}
\usepackage{soul}
\usepackage{amsfonts}

\usepackage{tikz}
\usetikzlibrary{arrows}
\usetikzlibrary{quantikz}

\usepackage[draft,inline,nomargin]{fixme} \fxsetup{theme=color}
\FXRegisterAuthor{cp}{acp}{\color{blue}CP}
\FXRegisterAuthor{tb}{ttb}{\color{green}TB}

\fancyheadoffset[L]{2.25in}
\fancyfootoffset[L]{2.25in}
\DeclareMathOperator{\tr}{tr}
\newtheorem{definition}{Definition}
\newtheorem{theorem}{Theorem}

\begin{document}
\vspace*{0.2in}

\begin{flushleft}
{\Large
\textbf\newline{Quantum simulation of Pauli channels and dynamical maps: algorithm and implementation } 
}
\newline
\\
Tomás Basile\textsuperscript{1,2\Yinyang},
Carlos Pineda\textsuperscript{2*\Yinyang},
\\
\bigskip
\textbf{1} Facultad de Ciencias. Universidad Nacional Autónoma de México, Ciudad de México 01000, Mexico
\\
\textbf{2} Instituto de Física, Universidad Nacional Autónoma de México, Ciudad de México 01000, México
\\
\bigskip

%
%
\Yinyang These authors contributed equally to this work.





* carlospgmat03@gmail.com

\end{flushleft}
\section*{Abstract}
Pauli channels are fundamental in the context of quantum computing as they
model the simplest kind of noise in quantum devices.  We propose a quantum
algorithm for simulating
Pauli channels 
and extend it to encompass Pauli dynamical maps 
(parametrized Pauli channels).
A parametrized quantum circuit is employed to accommodate for dynamical maps. 
We also establish the mathematical conditions for an
$N$-qubit transformation to be achievable using a parametrized circuit where
only one single-qubit operation depends on the parameter. 
The implementation of the proposed circuit is demonstrated using IBM's quantum computers 
for the case of one qubit, and the fidelity of this implementation is reported. 

\section{Introduction} 

Since their inception, quantum computers were proposed as powerful tools for
the simulation of quantum systems~\cite{feynman1982simulating}.  Being open
quantum systems of fundamental~\cite{Zur91,RevModPhys.76.1267} and
practical~\cite{breuer2007theory} interest, there has been efforts towards the
simulation of the evolution of open quantum systems~\cite{Garcia, Wang,Weimer}
and specifically for quantum channels~\cite{Xin,Wei,Zanetti}.

Such systems have been simulated because of their many applications, 
such as studying the emergence of multipartite entanglement~\cite{Andrea,Andrea_AD},
studying dissipative processes~\cite{Barreiro}
and modeling non-Markovian dynamics~\cite{Marsden}.
Among quantum systems, the simplest case is that of a qubit~\cite{chuangbook},
and withing them, the simplest class of channels that produce decoherence are
Pauli channels~\cite{geometry,Zbigniew,Davalos}. Indeed, they serve as
effective models for the noise affecting quantum devices~\cite{Flammia}.

To represent the algorithms implemented in quantum computers, either to
simulate a physical system or for some other purpose, one often uses a quantum
circuit~\cite{chuangbook}.  In this work, we shall also work with parametrized
quantum circuits, that is,  quantum circuits in which some of the operations
depend on variable parameters~\cite{cerezo}.  These circuits play an important
part in applications such as quantum machine learning~\cite{Benedetti} and
describing general quantum transformations dependent on parameters. Substantial
research has been devoted to enhancing the efficiency of these
circuits~\cite{Rasmussen}. 

We start by providing the definition of quantum channels, the general framework
used here, and multi-qubit Pauli channels in section \ref{sec: Pauli Channels}.
Our first objective is to present a quantum algorithm capable of simulating
Pauli channels on quantum computers; we do this in section \ref{sec: Circuit
for a Pauli Channel}, where we also demonstrate its implementation using IBM's
quantum computers for the particular case of single-qubit Pauli channels. 
Expanding beyond discrete Pauli channels, we introduce the concept of Pauli
dynamical maps, defined as a continuous parametrization of multi-qubit Pauli
channels.  Therefore, in section \ref{sec: 1PR Circuits} we shift our focus to
study parametrized quantum channels, aiming to adapt the algorithm developed
for channels to dynamical maps.  Furthermore, we contribute to the body of work
related to parametrized quantum circuits by establishing a theorem,
which  sets the mathematical conditions for the transformations
that can be done using a parametrized circuit with the condition that only a
controlled single-qubit
rotation in the circuit may depend on the parameter.  Finally, in
section \ref{sec: 1PR circuit for a Pauli map}, we conclude about the Pauli
dynamical maps that fulfill the conditions of theorem \ref{theorem2}. 

\section{Pauli channels and dynamical maps}  \label{sec: Pauli Channels} 


In this section  we introduce the concept of quantum channels, focusing on a
specific type called Pauli channels.  Furthermore, we define Pauli dynamical
maps, which are curves of Pauli channels parametrized by a variable.
\subsection{Quantum channels} \label{subsec: Quantum Channels} 

In quantum mechanics, a closed system's state is represented by a vector in a
Hilbert space $\mathcal{H}$.  The state's evolution is unitary and given by
Schrodinger's equation~\cite{Rieffel}.  However, in real-world situations,
quantum systems are usually open, which means that they interact with their
environment~\cite{breuer2007theory}.  For instance, the system's state may become
entangled with the environment, leading to a loss of information about the
system's state over time.

To describe open systems, instead of state vectors, we use matrices $\rho$ that
act on $\mathcal{H}$.  These matrices are called density matrices, and they
include information about the system's interaction with its environment.  For a
density matrix $\rho$ to be physically valid, it must satisfy two conditions:
$\tr(\rho) = 1$ and it must be positive semi-definite, which is denoted as
$\rho \geq 0$~\cite{chuangbook}.

Knowing this, we can now define quantum channels.  Quantum channels are
operators $\mathcal{E}$ that can describe the evolution of open quantum
systems, such that $\rho \rightarrow \mathcal{E}(\rho)$.  Quantum channels are
the most general linear operations that a quantum system can undergo
independently of its past~\cite{zimansbook,cirac}.  These channels are
constructed based on three fundamental properties: linearity, trace
preservation, and complete positivity.

Linearity ensures that a quantum channel $\mathcal{E}$ maps any ensemble of
density matrices into the corresponding ensemble of their evolution.  The trace
preserving property is given by $\tr (\mathcal{E}[\rho]) = \tr (\rho) = 1$ and
guarantees that the quantum channel does not change the condition that
$\tr(\rho) = 1$.  Finally, the channel should also preserve the condition $\rho
\geq 0$, and a map that does this is called a positive map.  However,
positivity of $\mathcal{E}$ is not enough, and we actually require the more
restrictive condition of complete positivity.  Complete positivity means that
$\mathcal{E} \otimes \mathbb{I}_n$ is positive for any positive integer $n$
(where $\mathbb{I}_n$ is the $n\times n$ identity matrix).  This ensures that
even if the principal system is entangled with another system, applying
$\mathcal{E}$ to the principal system while doing nothing to the other one
still results in  a positive semidefinite state for the principal
system~\cite{chuangbook}. 

Given a quantum channel $\mathcal{E}$, the condition of trace preservation is
straightforward to verify but complete positivity is not as simple.  To test
complete positivity of a quantum channel, Jamiołkowski and
Choi~\cite{choi,jamil} developed a simple algorithm that exploits the
isomorphism between a channel $\mathcal{E}$ and the state $\mathcal{D} =
(\mathbb{I} \otimes \mathcal{E}) [|\Omega \rangle \langle  \Omega|]$, where
$|\Omega\rangle = 1/\text{dim}(\mathcal{H}) \sum_{i}^{\text{dim}(\mathcal{H})}
|i \rangle |i \rangle$ is a maximally entangled state between the original
system and an ancilla.  Remarkably, the map $\mathcal{E}$ is completely
positive if and only if $\mathcal{D}$ (also known as the Choi or dynamical
matrix of $\mathcal{E}$) is positive semidefinite.

\subsection{Pauli channels}  \label{subsec: Pauli Channels} 

We have discussed the main features of quantum channels and 
now we turn our attention to a specific 
type of channels for $N$-qubit systems called Pauli channels.
First we will define these channels for single-qubit systems,
whose most general density matrix can be written as~\cite{chuangbook}:
\begin{equation}
\label{eq: Density Matrix}
\rho = \dfrac{1}{2} \sum_{\alpha=0}^{3} r_{\alpha} \sigma_{\alpha},
\end{equation}
with $\sigma_0 = \mathbb{I}$, and $\sigma_{1,2,3}$ the usual Pauli matrices.
The condition $\tr(\rho) = 1$ requires that $r_0 = 1$ while $\rho \geq 0$ 
implies that the remaining $r_{1,2,3}$ form
a vector $\vec{r}= (r_1,r_2,r_3)$ inside a unit sphere known as the Bloch sphere~\cite{Marinescu}.
That is, every possible density matrix for a one-qubit
system is uniquely associated with a point in a unit sphere. 

Given a one-qubit system described by $\rho$, 
a Pauli channel is defined as an operation that with probability $k_{\gamma}$
applies the Pauli matrix $\sigma_{\gamma}$ to
the system, for $\gamma = 0,1,2,3$~\cite{geometry}.
Mathematically, the Pauli channel is written in the following way:
\begin{eqnarray}
\label{eq: Pauli channel 1 qbit}
\mathcal{E}(\rho) = \sum_{\gamma=0}^3 k_{\gamma} \sigma_{\gamma} \rho \sigma_{\gamma},
\end{eqnarray} 
where the probabilities $k_{\gamma}$ of applying $\sigma_{\gamma}$
are non-negative real numbers such that 
$\sum_{\gamma} k_{\gamma} = 1$ (these conditions also ensure that the channel is 
trace preserving and completely positive).

Pauli channels are some of the 
most fundamental noise models in quantum information science~\cite{Terhal}. 
Some notable examples of Pauli channels are the following:
\begin{itemize}
\item \textbf{Bit Flip Channel:} This is a channel that with probability $1-p$ leaves the qubit as it is 
and with probability $p$ applies the $\sigma_1$ matrix 
(which flips the basis states $|0\rangle$ and $|1\rangle$ of the qubit), and
so it is given by:
\begin{align*}
\mathcal{E}(\rho) = (1-p) \rho + p \sigma_1 \rho \sigma_1.
\end{align*}
Analogous channels exist using $\sigma_3$ (called the bit flip channel,
which has a probability $p$ of adding a relative phase $\pi$ to the state) or 
using $\sigma_2$ (called the phase flip channel, 
which has a probability $p$ of 
flipping the base states and also add a relative phase $\pi$).
\item \textbf{Depolarizing channel:} 
This channel has a probability $1-p$
of doing nothing to the qubit and a probability $p$ of converting it into the maximally mixed state $\dfrac{1}{2} \mathbb{I}$
and it can be written as:
\begin{equation}
\mathcal{E}(\rho) = (1-p)\rho + p \dfrac{1}{2} \mathbb{I} = \left(1 - \dfrac{3p}{4} \right) \sigma + \dfrac{p}{4} \sigma_1 \rho \sigma_1 + \dfrac{p}{4} \sigma_2 \rho \sigma_2+ \dfrac{p}{4} \sigma_3 \rho \sigma_3.
\end{equation} 
\end{itemize}

We can also see how an arbitrary Pauli channel acts on an arbitrary  
density matrix.
To do it, we substitute \eref{eq: Density Matrix} 
in \eref{eq: Pauli channel 1 qbit}:
\begin{equation}
\mathcal{E}(\rho) = \dfrac{1}{2}\sum_{\gamma,\alpha=0}^3 k_{\gamma} r_{\alpha} \sigma_{\gamma} \sigma_{\alpha} \sigma_{\gamma}.
\end{equation}
This can be simplified by using the following property of Pauli matrices:
\begin{equation}
\label{eq: propiedad-pauli}
\sigma_{\gamma} \sigma_{\alpha} \sigma_{\gamma} = A_{\alpha,\gamma} \sigma_{\alpha}, \;\;\;\;\; \text{with} \; A_{\alpha,\gamma} = \begin{pmatrix}
1 & 1 & 1 & 1\\
1 & 1 & -1 &-1 \\
1 & -1 & 1 & -1 \\
1 & -1 & -1 & 1
\end{pmatrix},
\end{equation}
which leads to
\begin{equation}
\label{eq: rho-transformada}
\mathcal{E}(\rho) = \dfrac{1}{2} \sum_{\alpha} \left(\sum_{\gamma} A_{\alpha, \gamma} k_{\gamma} \right) r_{\alpha} \sigma_{\alpha}.
\end{equation}
\eref{eq: rho-transformada} once again has the form of \eref{eq: Density Matrix}
but with components $\left( \sum_{\gamma} A_{\alpha,\gamma} k_{\gamma} \right) r_{\alpha}$.
This gives us another way of understanding Pauli channels
as operations that take each component $r_{\alpha}$
of the density matrix and multiplies them
by $\sum_{\gamma} A_{\alpha,\gamma} k_{\gamma}$, 
that is:
\begin{equation}
\label{eq: multipliers}
r_{\alpha}  \xrightarrow[\text{Channel}]{\text{Pauli}}  \tau_{\alpha} r_{\alpha} ,\;\; \tau_{\alpha} := \sum_{\gamma} A_{\alpha,\gamma} k_{\gamma}.
\end{equation}
Notice that $\tau_0 = 1$, which is a consequence of $\sum_{\gamma}k_{\gamma}=1$
and ensures that after the channel, the resulting density matrix still has trace one. 
Furthermore, reverting the definition of $\tau_{\alpha}$ by using that $A^{-1} = \dfrac{1}{4} A$, 
we get that $k_{\gamma} = \dfrac{1}{4} \sum_{\alpha} A_{\alpha,\gamma} \tau_{\alpha}$.
Then, using that $k_{\gamma} \geq 0$
we get the following conditions on the multipliers $\tau_{\alpha}$:
\begin{eqnarray}
\label{eq: conditions-tetrahedron}
&1+\tau_i -\tau_j - \tau_k \geq 0,  \;\;\; \text{for $i,j,k$ different numbers in $\{1,2,3\}$}, \\
&1+\tau_1 + \tau_2 + \tau_3 \geq 0.
\end{eqnarray}

These conditions imply that $(\tau_1,\tau_2,\tau_3)$
has to be inside a tetrahedron with
vertices $(1,1,1), (1,-1,-1), (-1,1,-1)$ and $(-1,-1,1)$. 
Therefore, the $\tau_{1,2,3}$ are numbers between 
$-1$ and $1$, which means that the components
$r_{\alpha}$ of the density matrix are always 
attenuated and possibly sign flipped.

Having defined the one qubit case,  we can now generalize to $N$ qubits.
In order to do it, we need to introduce the so-called \textit{Pauli strings}, defined as
\begin{eqnarray}
\label{eq: Pauli string}
\sigma_{\vec{\alpha}} = \sigma_{\alpha_1} \otimes \sigma_{\alpha_2}\otimes \cdots \otimes \sigma_{\alpha_N},
\end{eqnarray}
where $\vec{\alpha}$ denotes a multi-index $(\alpha_1, \cdots, \alpha_N)$
 and $\alpha_i \in \{0,1,2,3\}$. 
These operators form an orthogonal basis in the space of operators acting on $N$ qubits. 
 Similarly to the single-qubit case, the density matrix $\rho$ 
of a system of $N$ qubits can be written using Pauli strings as:
\begin{equation}
\label{eq:  Density matrix Nqbit}
\rho = \dfrac{1}{2^N} \sum_{\vec{\alpha}} r_{\vec{\alpha}} \sigma_{\vec{\alpha}}.
\end{equation}
Then, just as before, we define a Pauli channel as a transformation that applies 
the operator $\sigma_{\vec{\gamma}}$ to $\rho$ with probability $k_{\vec{\gamma}}$
and is therefore described mathematically by:
\begin{equation}
\label{eq: pauli channel N-qubit}
\mathcal{E}(\rho) = \sum_{\vec{\gamma}} k_{\vec{\gamma}} \sigma_{\vec{\gamma}} \rho \sigma_{\vec{\gamma}},
\end{equation}
where just as before, $k_{\vec{\gamma}}$ are non-negative real numbers such 
that $\sum_{\vec{\gamma}} k_{\vec{\gamma}}=1$.

As in the one qubit case, Pauli channels for $N$ qubits 
attenuate the components $r_{\vec{\alpha}}$ of the density matrix.
This can be seen by substituting \eref{eq:  Density matrix Nqbit} in \eref{eq: pauli channel N-qubit}
and using the property of \eref{eq: propiedad-pauli}:
\begin{align*}
\mathcal{E}(\rho) &= \dfrac{1}{2^N} \sum_{\vec{\gamma},\vec{\alpha}} k_{\vec{\gamma}} r_{\vec{\alpha}} \sigma_{\vec{\gamma}} \sigma_{\vec{\alpha}} \sigma_{\vec{\gamma}} \\
& = \dfrac{1}{2^N} \sum_{\vec{\alpha}} \left( \sum_{\vec{\gamma}} {(A^{\otimes^N})}_{\vec{\alpha},\vec{\gamma}} \; k_{\vec{\gamma}} \right) r_{\vec{\alpha}} \sigma_{\vec{\alpha}},
\end{align*}
which means that applying the Pauli channel multiplies the 
components $r_{\vec{\alpha}}$ by $\tau_{\vec{\alpha}} := \sum_{\vec{\gamma}} {(A^{\otimes^N})}_{\vec{\alpha},\vec{\gamma}} k_{\vec{\gamma}}$.
\subsection{Pauli dynamical maps} 
\label{subsec: Pauli Dynamical Maps}

As seen in the last section, Pauli channels and in 
general quantum channels are discrete maps
that transform a density matrix $\rho$ into $\mathcal{E}(\rho)$.
However, we could also define a continuous set of 
channels $\varepsilon_p$ with $p$ a real parameter.

For the special case of Pauli channels, we define 
a Pauli dynamical map as a continuous parametrized 
curve drawn inside the set of Pauli channels and starting at the identity channel. 
Therefore, a Pauli dynamical map can be written as
\begin{eqnarray}
\label{eq: Pauli dynamical map}
\mathcal{E}_p(\rho) = \sum_{\vec{\gamma}} k_{\vec{\gamma}}(p) \sigma_{\vec{\gamma}} \rho \sigma_{\vec{\gamma}},
\end{eqnarray}
where $p$ is a parameter in an interval $[a,b]$ 
and $\mathcal{E}_p$ is a Pauli channel for every $p$, 
with $\mathcal{E}_a$ being the identity channel.

\section{Circuit for a Pauli channel} 
\label{sec: Circuit for a Pauli Channel}

In this section we propose a quantum circuit that simulates $N$-qubit Pauli
channels.  Moreover, we implement the circuit for $N=1$ on a quantum computer
and analyze the results using the diamond norm.
We find that close to the depolarizing channel, the general circuit simulator 
can implement such channels with the highest fidelity. 
\subsection{Description of the circuit for a Pauli channel} 
\label{subsec: Description of the circuit}

\begin{figure} 
\centering
\includegraphics{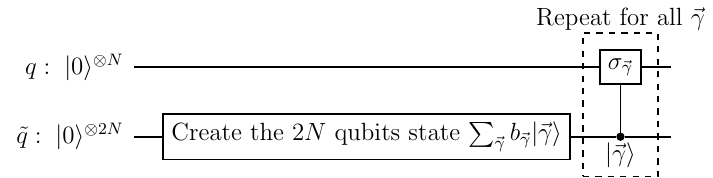}
\caption{{\bf Circuit for an $N$-qubit
Pauli channel.}
The circuit creates the state $\sum_{\vec{\gamma}} b_{\vec{\gamma}}|\vec{\gamma}\rangle$ 
on the $2N$ ancilla qubits denoted by $\tilde{q}$.}
\label{Fig1}
\end{figure}

To design the circuit that implements \eref{eq: pauli channel N-qubit}, we construct a state that 
includes the probabilities $k_{\gamma}$ on the ancilla qubits and subsequently
apply controlled Pauli operations on the main qubits.  The circuit that does
this is presented in \fref{Fig1}.

The first part of the circuit involves the creation of the state
\begin{eqnarray}
\label{eq: state}
\text{Ancilla state =} 
\sum_{\vec{\gamma}} b_{\vec{\gamma}} |\vec{\gamma} \rangle
\end{eqnarray}
on the ancilla qubits, where $b_{\vec{\gamma}}$ are numbers such that $
{|b_{\vec{\gamma}}|}^2 = k_{\vec{\gamma}}$ and
the $2N$-qubit state $|\vec{\gamma}\rangle$ is defined as $|\gamma_1\rangle
\cdots |\gamma_N\rangle$.
When measured in the computational basis, the state given in \eref{eq: state}
collapses to $|\vec{\gamma}\rangle$ with a 
probability ${|b_{\vec{\gamma}}|}^2 = k_{\vec\gamma}$. 
The circuit in \fref{Fig1} uses this fact
to apply $\sigma_{\vec{\gamma}}$ on the main qubits
with a probability $k_{\vec\gamma}$ by using controlled operations 
conditioned on the state of the system being $|\vec{\gamma}\rangle$,
just as the Pauli channel is supposed to do.  

\subsection{Simulation for one-qubit Pauli channels} 
\label{subsec: Simulation for one-qubit Pauli channels}

\begin{figure} 
\centering
\includegraphics{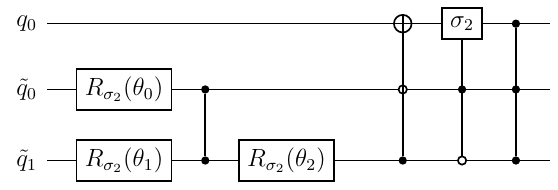}
\caption{
{\bf One-qubit Pauli channel circuit.} Circuit for a one-qubit Pauli channel,
which is a particular case of \fref{Fig1}.  Here we have two
ancilla qubits and we use three rotations
of angles given by \eref{eq:angle} to create the Ancilla State of
of \eref{eq: state} for the two ancilla qubits.
}
\label{Fig2} 
\end{figure}

For the particular case of a Pauli channel on one qubit, the circuit that
simulates it can be constructed as in \fref{Fig2},
which is a special case of \fref{Fig1}
but with all details  explicitly shown.
In said figure, the ancilla state
of \eref{eq: state} can be taken to be
$\sqrt{k_{0}} |00\rangle + \sqrt{k_{1}} |01\rangle + \sqrt{k_{2}} 
|10\rangle + \sqrt{k_{3}}|11\rangle$
and it is created on the ancilla qubits with the help of three rotations of angles
defined by the following equations:
\begin{align}
\label{eq:angle}
\cos\left(\dfrac{\theta_0}{2} \right) &= \sqrt{k_{0} + k_{1}},\nonumber \\
\tan\left( \dfrac{\theta_1 + \theta_2}{2} \right) &= \sqrt{k_{1}/k_{0}},  \\
\tan\left( \dfrac{\theta_2 - \theta_1}{2} \right) &= \sqrt{k_{3}/k_{2}}.  \nonumber
\end{align}

\begin{figure} 
\centering
\includegraphics{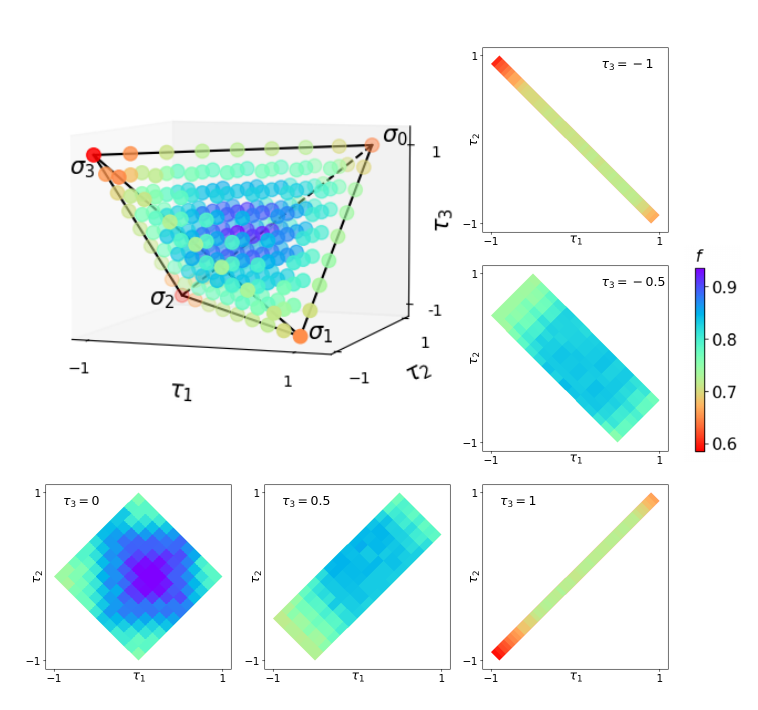}
\caption{{\bf Results of the diamond fidelities as defined in \eref{eq:diamond-fid}
for a sample of Pauli channels in the tetrahedron.}  Notice that channels close
to the center of the tetrahedron have high fidelities, while those close to the
borders do not.  Moreover, we show the results for cuts of the tetrahedron at
different values of $\tau_3$.}
\label{Fig3}
\end{figure}

We took a sample of one-qubit Pauli channels
and evaluated their implementation on IBM's
ibmq-lima quantum computer~\cite{Qiskit},
as shown in \fref{Fig3}. 
For each of the channels sampled, we used 
quantum process tomography~\cite{Qiskit, Chuang:1996} to obtain the 
operator $\xi_I$ corresponding to the implementation of the
circuit in the quantum computer. 
Then, we compared $\xi_I$ with the theoretical 
operator $\xi_T$ of the Pauli channel we wanted to implement.
To see how close the operators $\xi_I$ and 
$\xi_T$ are, we shall use the
diamond distance~\cite{wildebook}, which is defined by
\begin{equation}
||\xi_I - \xi_T ||_{\diamond}  = \max_{\rho} || (\xi_I \otimes I) \rho - (\xi_T \otimes I) \rho ||_1,
\end{equation}
with $I$ the identity map,
$|| \cdot ||_1$ the trace norm and the maximization
done over all density matrices $\rho$.
The calculation of this norm is done using the semi-definite program from reference \cite{Watrous}.
When the two channels are the same,
the diamond distance has a value of $0$, while in the case that the channels
are completely distinguishable, the distance reaches its maximum value of $2$~\cite{Benenti}.
For the analysis done in \fref{Fig3}, we define a sort of ``diamond fidelity'' as: 
\begin{equation}
f = 1 - \dfrac{1}{2} ||\varepsilon_I - \varepsilon_T ||_{\diamond},
\label{eq:diamond-fid}
\end{equation}
which ranges from $0$, when the channels have a maximum distance, to  $1$,
when they are exactly equal.

Finally, using the representation of Pauli channels in a tetrahedron as in
\eref{eq: conditions-tetrahedron}, we show in \fref{Fig3} the diamond
fidelity defined by \eref{eq:diamond-fid} for the channels analyzed.  We can
see that channels close to the completely depolarizing channel (that is close
to the center of the tetrahedron) have a high $f$, while those close to unitary
channels have much lower $f$.  This is reasonable because quantum computers are
prone to errors that depolarize qubits, which isn't very problematic when
trying to simulate depolarization but it is when simulating unitary processes.
Moreover, the algorithm of \fref{Fig2} is not optimal for
unitary channels (that is, the channels corresponding to the vertices of the
tetrahedron).  These straightforward channels could be accomplished more
efficiently by simply applying the corresponding Pauli operation directly.
Nevertheless, due to its general design to accommodate any Pauli channel, the
algorithm employs numerous quantum gates even in such scenarios.

\section{One parameter circuits} 
\label{sec: 1PR Circuits}

Just as Pauli channels, Pauli dynamical maps can be implemented using the circuit
of \fref{Fig1}. However, there is one difference: 
the state to be created on the ancilla qubits now 
depends on a parameter $p$, and it is represented by the expression:
\begin{eqnarray}
\label{eq: parametrized state}
\sum_{\vec{\gamma}} b_{\vec{\gamma}}(p) |\vec{\gamma}\rangle.
\end{eqnarray}
Thus, we temporarily shift our focus from Pauli
channels and dynamical maps to the general problem of 
creating a circuit to generate a curve of
states like the one described in \eref{eq: parametrized state}.

In general, producing this curve of states for $N$ qubits will require
many rotations parametrized by $p$,
such as the three rotations used for the ancilla
qubits in \fref{Fig2}.
However, it would be preferable to achieve the same effect using only one parametrized rotation.
This would allow us to interpret said rotation 
as a knob that smoothly traverses the curve of states.
Consequently, we are faced with the question of which curves of states, 
such as the one described in \eref{eq: parametrized state}, 
can be produced using just a single parametrized rotation. 
To clarify this, we provide the following definition 
for a circuit with one parametrized rotation.

\begin{definition}{\textbf{1-Parameter Rotation Circuit:}}
A 1-Parameter Rotation (1PR) circuit is a parametrized quantum
circuit that includes only one gate dependent on a parameter $p$.
Moreover, the parametrized gate is a one-qubit rotation about any axis,
whether controlled or not.
\end{definition}

\begin{figure} 
\centering
\includegraphics{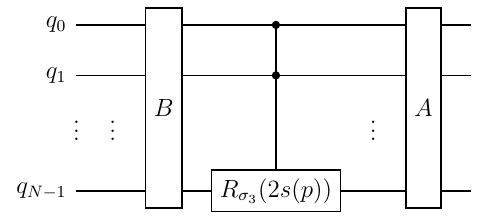}
\caption{\textbf{General form of a 1PR circuit}.
Any 1PR circuit can be transformed into this form,
where the rotation on the last qubit
can be controlled or not by any of the other qubits.
$A$ and $B$
are $N$-qubit gates that do not depend on the parameter $p$ and
$s= s(p)$ is a function of the parameter.}
\label{Fig4}
\end{figure}

Based on this definition, we aim to determine which curves of 
states can be generated using 1PR circuits. 
To accomplish this, we begin by proving that all 1PR circuits have 
the form depicted in \fref{Fig4},
where the parametrized rotation is around $\sigma_3$ and
is applied to the last qubit.

\begin{theorem}
An $N-$qubit 1PR circuit can always be transformed 
into the form shown in \fref{Fig4}.
\end{theorem}
\textbf{Proof:} 
First, we observe that according to the
definition, a 1PR circuit always consists of an operation $B$ followed 
by the parametrized rotation and then another operation $A$, 
where $A$ and $B$ are not parametrized.

Next, we note that it is not necessary to consider rotations about an arbitrary axis,
as a rotation about any axis $\hat{n}$ parameterized by $p$
can be transformed into a rotation about $\sigma_3$ without introducing 
gates that depend on $p$. To see this, consider the rotation $R_{\hat{n}}(2s)$, where $2s$ 
is a function of $p$ (the factor of 2 is for convenience later on) 
and $\hat{n} = (n_1,n_2,n_3)$ represents the rotation axis. 
We can express $\hat{n}$ as $(\sin \theta \cos \phi, \sin \theta \sin \phi, \cos \theta)$, 
where $\theta$ and $\phi$ are fixed angles dependent on $\hat{n}$. 
The rotation can then be rewritten as follows:
\begin{eqnarray}
R_{\hat{n}}(2s) = R_{\sigma_3}(\phi) R_{\sigma_2}(\theta) R_{\sigma_3}(2s) R_{\sigma_2}(-\theta) R_{\sigma_3}(-\phi).
\end{eqnarray}
Since the angles $\theta$ and $\phi$ do not depend on the parameter $p$,
any 1PR circuit can be transformed into a circuit where the parametrized
 rotation is around $\sigma_3$ instead of an arbitrary axis. 
Moreover, without loss of generality, we can choose the last qubit as the
target qubit for the rotation, since if it weren't, we could use
swap gates to move the rotation
to the first qubit without adding gates
that depend on $p$. 

Therefore, a 1PR circuit can be transformed such
that the rotation is around $\sigma_3$ and is applied to the last qubit 
(possibly controlled by other qubits), 
resulting in the form depicted in \fref{Fig4}. 
$\blacksquare$ \\
$\;$\\

With the aid of this theorem, we can now determine the curves
of states of $N$ qubits that can be generated using a 1PR circuit. 
This result is stated in the following theorem.

\begin{theorem}
\label{theorem2}
Consider a 1PR circuit of $N$ qubits parametrized by $p$
and denote by $U$ the operator it implements on this system. 
Then, for every $j \in \{0, 1, \cdots, 2^N-1\}$, 
we have that:
\begin{align*}
U|j\rangle = e^{is(p)} |a^j\rangle + e^{-is(p)} |b^j\rangle + |c^j\rangle
\end{align*}
with $s(p)$ some function of $p$,  $|a^j\rangle ,|b^j\rangle, |c^j\rangle$ orthogonal states and $\langle a^j| a^j\rangle + \langle b^j| b^j\rangle + \langle
c^j|c^j \rangle = 1$.
\end{theorem}
\textbf{Proof:} 
We can conclude from theorem 1 that $U=ARB$,
where $A$ and $B$ are unitary matrices and $R$ is a $\sigma_3$ rotation of angle $2s$
applied to the last qubit and controlled by some of the other ones. 

First, applying $B$ to $|j\rangle$ results in $B|j\rangle = B_{0,j} |0\rangle + B_{1,j} |1 \rangle + \cdots + B_{2^n-1,j}|2^n-1\rangle$,
with $B_{i,j}$ the entries of matrix $B$.
This can be rewritten by separating last qubit from the other $N-1$:
\begin{equation}
B|j\rangle = \sum_{k=0}^{2^{N-1}-1} \left( B_{2k,j} |k\rangle|0\rangle  + B_{2k+1,j} |k \rangle |1\rangle  \right).
\end{equation}
After the operator $B$, the circuit applies the  controlled rotation $R$.
To simplify the analysis, we separate the states of
the first $N-1$ qubits into those that fulfill the control conditions of the rotation
(which we denote as the set $\mathcal{C}$)
and those that do not, and write it as
\begin{align}
B|j\rangle = 
   \sum_{k \in \mathcal{C}} \left( B_{2k,j} |k\rangle|0\rangle 
                                     + B_{2k+1,j} |k \rangle |1\rangle  \right) 
   + \sum_{k \not\in \mathcal{C}} \left( B_{2k,j} |k\rangle|0\rangle  
                                      + B_{2k+1,j} |k \rangle |1\rangle  \right).
\end{align}

Then, the rotation $R$ will only affect the states on the first sum (since they fulfill the control conditions)
and not the others. Therefore,  
remembering that a $\sigma_3$ rotation acts by adding a phase $e^{-is(p)}$ to $|0\rangle$
and a phase $e^{is(p)}$ to $|1\rangle$, we have that,
\begin{align}
RB|j\rangle& = e^{-is(p)} \sum_{k \in \mathcal{C}} B_{2k,j} |k\rangle |0\rangle + e^{is(p)} \sum_{k \in \mathcal{C}} B_{2k+1,j} |k\rangle |1\rangle 
+ \sum_{k \not\in \mathcal{C}} \left( B_{2k,j} |k\rangle|0\rangle  + B_{2k+1,j} |k \rangle |1\rangle  \right) \nonumber \\
& = e^{-is(p)} |\tilde{b}^{j}\rangle + e^{is(p)} |\tilde{a}^j\rangle + |\tilde{c}^j\rangle,
\end{align}
where we defined
\begin{align*}
|\tilde{a}^j \rangle &= \sum_{k \in \mathcal{C}} B_{2k,j}|k\rangle|0\rangle,\\
|\tilde{b}^j\rangle &= \sum_{k \in \mathcal{C}} B_{2k+1,j} |k\rangle |0 \rangle,\\
|\tilde{c}^j\rangle &= \sum_{k \not\in \mathcal{C}}\left( B_{2k,j} |k\rangle|0\rangle+ B_{2k+1,j} |k \rangle |1\rangle  \right).
\end{align*}
These states are clearly orthogonal because they are each linear combinations of different 
orthogonal states of the computational basis. 
Moreover, they satisfy  $\langle a^j| a^j\rangle + \langle b^j| b^j\rangle + \langle c^j| c^j\rangle = 1$ because this quantity is the squared norm of the $j$th column of $B$, 
which is unitary.

Finally, after having applied the rotation, the circuit applies gate $A$, 
so that the result is given by:
\begin{eqnarray}
U|j\rangle = ARB|j\rangle = e^{-is(p)} A |\tilde{a}^j\rangle + e^{is(p)} A |\tilde{b}^j\rangle + A |\tilde{c}^j\rangle = e^{-is(p)} |a^j\rangle + e^{is(p)} |b^j\rangle + |c^j\rangle,
\end{eqnarray}
where $|a^j\rangle = A |\tilde{a}^j\rangle, |b^j\rangle = A |\tilde{b}^j\rangle, |c^j\rangle = A |\tilde{c}^j\rangle$ are still orthogonal
states that satisfy $\langle a| a\rangle + \langle b| b\rangle + \langle c| c\rangle = 1$ because $A$ is unitary. $\blacksquare$  \\
 $\;$ \\

This theorem implies that when starting from the state $|0\rangle$ 
or any other initial state, the only possible curves of states that can be 
created using a 1PR circuit are of the following form:
\begin{equation}
|\eta(p)\rangle = |c\rangle + e^{is(p)}|a\rangle + e^{-is(p)} |b\rangle,
\label{eq:curve-states}
\end{equation}
with conditions defined by the equations: 
\begin{equation}
\langle a| a\rangle + \langle b| b\rangle + \langle c| c\rangle = 1,  \quad
\langle a |b\rangle =
\langle a |c\rangle =
\langle b |c\rangle = 0.
\label{eq:conditions-vecs}
\end{equation}

Moreover, it is possible to construct a 1PR circuit to generate any given curve of 
states described by \eref{eq:curve-states}. 
One approach to achieve this is by utilizing the circuit 
depicted in \fref{Fig4}, 
with the parametrized rotation $R$ applied to the last qubit controlled by all the 
other qubits. The operators $A$ and $B$ can be defined as follows:
\begin{align*}
&B|0\rangle = \sqrt{\langle a | a \rangle} |2^{N}-1 \rangle + \sqrt{\langle b | b \rangle} |2^{N}-2\rangle + \sqrt{\langle c |c \rangle} |2^N-3\rangle, \\
&A|2^N-1 \rangle = \frac{1}{\sqrt{\langle a | a \rangle}} |a\rangle,\; A|2^{N}-2\rangle = \frac{1}{\sqrt{\langle b | b \rangle}}|b\rangle, \; A|2^{N}-3\rangle = \frac{1}{\sqrt{\langle c | c \rangle}}|c\rangle.
\end{align*}
The remaining part of the operators $A$ and $B$ can be defined
in any arbitrary manner as long as they are unitary.
By starting from the initial state $|0\rangle$ and 
applying the circuit shown in \fref{Fig4}, straightforward calculations 
lead us to obtain the resulting curve of states described in \eref{eq:curve-states}.

To see it, we can rewrite the expression of $B|0\rangle$ by 
separating the first $N-1$ qubits from the last one:
$$B|0\rangle = \sqrt{\langle a | a \rangle} |2^{N-1}-1\rangle |1\rangle + \sqrt{\langle b | b \rangle} |2^{N-1}-1\rangle |0\rangle + \sqrt{\langle c | c \rangle} |2^{N-1}-2\rangle |1\rangle.$$
Since the parametrized rotation $R$ is controlled by all the first $N-1$ qubits, it only applies to the first two terms of $B|0\rangle$. As a result, we obtain:

$$
RB|0\rangle =\sqrt{\langle a | a \rangle} e^{is(p)} |2^{N-1}-1\rangle |1\rangle + \sqrt{\langle b | b \rangle} e^{-is(p)} |2^{N-1}-1\rangle |0\rangle + \sqrt{\langle c | c \rangle} |2^{N-1}-2\rangle |1\rangle.
$$

Finally, applying the defined operator $A$ to this state yields the desired result.
 
\section{1PR circuit for a Pauli map} 
\label{sec: 1PR circuit for a Pauli map}

We can now use the previous results to conclude directly which Pauli dynamical maps
can be implemented with a 1PR circuit.
For this, the curve of states of \eref{eq: parametrized state} 
has to be constructed with only one parametrized rotation,
so it has to satisfy the conditions of theorem \ref{theorem2}.
Therefore, this implies that the map
\begin{eqnarray}
\varepsilon_p(\rho) = \sum_{\vec{\gamma}} k_{\vec{\gamma}}(p) \sigma_{\vec{\gamma}} \rho \sigma_{\vec{\gamma}},
\end{eqnarray}
can be implemented  if there are numbers $\beta_{\vec{\gamma}}(p)$ such that $|\beta_{\vec{\gamma}}(p)|^2 = k_{\vec{\gamma}}(p)$ and
\begin{eqnarray}
\label{eq:vec}
\sum_{\vec{\gamma}} \beta_{\vec{\gamma}}(p) |\vec{\gamma}\rangle = |c\rangle +  e^{is(p)} |a\rangle + e^{-is(p)}|b\rangle,
\end{eqnarray}
where $|a\rangle,|b\rangle,|c\rangle$ fulfill
the conditions of \eref{eq:conditions-vecs}.

\begin{figure} 
\centering
\includegraphics{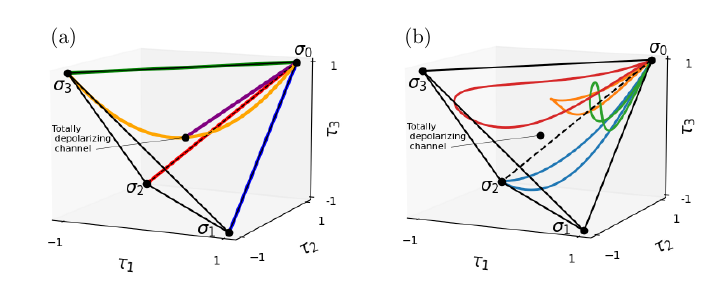}
\caption{{\bf Some Pauli dynamical maps that can be implemented with a 1PR circuit.}
The curves painted in these tetrahedrons
represent Pauli dynamical maps that can be implemented with 1PR circuits.
(a) shows the dynamical maps mentioned in the main 
text, which are: depolarizing (purple), bit flip (blue),
phase flip (green), bit-phase flip (red)
and parabolic (orange).
(b) shows dynamical maps selected at random that can be implemented with 1PR 
circuits.}
\label{Fig5}
\end{figure}

For the particular case of one qubit, we can show some examples of Pauli
dynamical maps implementable with a 1PR circuit, which are plotted in
\fref{Fig5}. 
The examples we show include some of the most common maps: the bit flip, phase flip, bit-phase flip and depolarizing.
However, we also include the parabolic dynamical map,
defined in \eref{eq:parab} and shown in \fref{Fig5}.
This map traces a parabola inside the tetrahedron connecting two of its vertices
and it describes a frontier in the tetrahedron between Pauli channels
that are reachable by Lindbladian dynamics and those that are not~\cite{Davalos}.

\begin{itemize}
\item \textbf{Depolarizing:} This dynamical map is given by
\begin{align*}
\varepsilon_p(\rho) = (1-3p/4) \rho + (p/4) \sigma_1 \rho \sigma_1 + (p/4) \sigma_2 \rho \sigma_2 + (p/4) \sigma_3 \rho \sigma_3,
\end{align*}
with $p \in [0,1]$. 
Therefore, the curve of states $|\beta(p)\rangle$ needed on the ancilla
qubits is such that ${|\beta_0(p)|}^2= (1-3p/4)$, ${|\beta_1(p)|}^2 =
{|\beta_2(p)|}^2 = {|\beta_3(p)|}^2 = p/4$.  Then, taking the $\beta_j$ to be
real, the curve of states can be 
\begin{align*}
|\beta(p) \rangle = \sqrt{1-3p/4} |00\rangle + \sqrt{p/4} |01\rangle + \sqrt{p/4} |10 \rangle  + \sqrt{p/4} |11\rangle.
\end{align*}
This state can be rewritten as:
\begin{align*}
|\beta(p)\rangle =& \; e^{is} \left( \dfrac{1}{2} |00\rangle - \dfrac{i}{2\sqrt{3}} |01\rangle 
      - \dfrac{i}{2\sqrt{3}} |10 \rangle - \dfrac{i}{2\sqrt{3}} |11\rangle \right) \\
& + e^{-is} \left( \dfrac{1}{2} |00\rangle + \dfrac{i}{2\sqrt{3}} |01\rangle 
      + \dfrac{i}{2\sqrt{3}} |10\rangle + \dfrac{i}{2\sqrt{3}} |11 \rangle \right),
\end{align*}
with $\sin s = \sqrt{3p/4}$.
We can see that this curve satisfies the conditions of \eref{eq:conditions-vecs}, meaning that it can be created with a 
1PR circuit.

\item \textbf{Parabolic dynamical map:} We define the parabolic dynamical map as:
\begin{eqnarray}
\epsilon(\rho) = \dfrac{1}{4} {(1-p)}^2  \rho + \dfrac{1}{4} (1-p^2) \sigma_1 \rho \sigma_1 + \dfrac{1}{4} (1-p^2) \sigma_2 \rho \sigma_2 + \dfrac{1}{4} {(1+p)}^2 \sigma_3 \rho \sigma_3,
\label{eq:parab}
\end{eqnarray}
with $p \in [-1,1]$. 
If we take the $\beta_j$ to be real,
the curve of states needed on the ancilla qubits can be:
\begin{eqnarray}
|\beta(p)\rangle = \dfrac{1}{2} (1-p) |00\rangle + \dfrac{1}{2} \sqrt{1-p^2} |01\rangle +  \dfrac{1}{2} \sqrt{1-p^2} |10\rangle +  \dfrac{1}{2}(1+p) |11\rangle.
\end{eqnarray}
This can be rewritten as
\begin{align*}
|\beta(p)\rangle =&\; \left( \dfrac{1}{2}|00\rangle + \dfrac{1}{2} |11\rangle \right) + e^{is} \left( \dfrac{i}{4}|00\rangle + \dfrac{1}{4} |01\rangle + \dfrac{1}{4} |10\rangle - \dfrac{i}{4} |11\rangle \right) \\
& + e^{-is}  \left( \dfrac{-i}{4} |00\rangle + \dfrac{1}{4} |01\rangle + \dfrac{1}{4} |10\rangle + \dfrac{i}{4} |11\rangle \right),
\end{align*}
with $\sin s = p$, so that
this map fulfills the conditions of  \eref{eq:conditions-vecs}.

\item \textbf{Bit flip map:} 
This dynamical map is defined as
\begin{align*}
\varepsilon_p(\rho) = (1-p)\rho + p\sigma_1 \rho \sigma_1,
\end{align*}
for $p \in [0,1]$. 
In particular, if we take the $\beta_j$ to be real,
we need to create the curve of states:
\begin{align*}
|\beta(p)\rangle = \sqrt{1-p} |00\rangle +\sqrt{p} |01\rangle.
\end{align*}
This can be rewritten as
\begin{align*}
|\beta(p) \rangle = e^{is}  \left(\frac{1}{2} |00\rangle - \dfrac{i}{2} |01\rangle \right) + e^{-is} \left( \frac{1}{2} |00\rangle + \frac{i}{2} |01\rangle \right),
\end{align*}
with $\sin s= \sqrt{p}$.
Therefore, we can see that the curve satisfies the conditions of  \eref{eq:conditions-vecs} 
and it can be created with a 1PR circuit.
Note that in this case we actually only need one ancilla qubit
since the state $|\beta(p)\rangle$ is only two dimensional.

The exact same thing can be done
for the phase flip and bit phase flip dynamical maps by
changing $\sigma_1$ to
$\sigma_3$ and $\sigma_2$ respectively.
For example, the bit phase flip map was implemented in \cite{Andrea, Andrea_4qb} 
using an optical arrangement, and  it was indeed done 
by varying only one angle that depends on the parameter $p$ (the angle of a half waveplate).

\end{itemize}

Furthermore, we can construct other examples of 
Pauli dynamical maps such that they
can be implemented with a 1PR circuit.
To do it, we only need to choose the three states $|a\rangle$, $|b\rangle$, $|c\rangle$
that satisfy the conditions of  \eref{eq:conditions-vecs}.
For example, this can be done systematically for the case of curves of states of two qubits
(that is, for Pauli dynamical maps of one qubit)
with the following procedure:
\begin{itemize}
\item[1.] We first choose the norms $|a|$, $|b|$, $|c|$
such that $\langle a| a\rangle + \langle b| b\rangle + \langle c| c\rangle = 1$. 
This can be done by selecting two angles
$\mu \in [0, \pi/2]$, $\nu \in [0,\pi/2]$ and defining:
\begin{align*}
|a| = \sin \nu \cos \mu, \; |b| = \sin \nu \sin \mu , \; |c| = \cos \nu.
\end{align*}
\item[2.] We define $|a'\rangle = |a| |0\rangle$, $|b'\rangle = |b| |1\rangle$, $|c'\rangle = |c| |2\rangle$.
\item[3.] 
Finally, we choose a unitary matrix $V$ with the condition that its first
row is equal to $e^{i\theta} (|a|,|b|,|c|,0)$
with $\theta$ a uniform random phase.
That way, we can define $|a\rangle = V |a'\rangle$, $|b\rangle = V |b'\rangle$, $|c\rangle = V |c'\rangle$
and since $V$ is unitary, these unprimed vectors will fulfill the conditions of
\eref{eq:conditions-vecs}.
Furthermore, the form of the first row ensures that the dynamical map begins at the identity,
since it implies that when $s=0$, the 
state created in \eref{eq:vec}
is
$|a\rangle + |b\rangle + |c\rangle = e^{i \theta} |0\rangle$,
which  corresponds with applying the identity channel.

Such a matrix $V$ can be randomly constructed by first finding three vectors 
$\vec{w}_1 , \vec{w}_2, \vec{w}_3$ orthogonal to
the first row using the Gram-Schmidt process.  Then selecting random complex
numbers $r_1, r_2, r_3$ such that $|r_1|^2 + |r_2|^2 + |r_3|^2 = 1$ and defining
the second row of $V$ to be $r_1 \vec{w}_1 + r_2 \vec{w}_2 + r_3 \vec{w}_3$.
Once the first two rows are chosen, use Gram-Schmidt to find two vectors
$\vec{v}_1, \vec{v}_2$ orthogonal to 
them and similarly define the third row as $q_1 \vec{v}_1+ q_2 \vec{v}_2$
with $|q_1|^2 + |q_2|^2$ and $q_1,q_2$
selected at random. 
Finally, there is only one choice for the fourth row so that it is 
orthonormal to the first three
and a random phase can be given to it.
\end{itemize}

Following this procedure for random
angles and unitary matrices $V$, 
we plot four Pauli dynamical maps  selected at random
that can be implemented with a 1PR circuit in \fref{Fig5}.

\section{Conclusion} 

In this work, we found a quantum algorithm for 
simulating Pauli channels in $N$-qubit systems and
generalized it to Pauli dynamical maps
by using parametrized quantum circuits.
Furthermore, we implemented single-qubit Pauli channels
on one of IBM's quantum computers
and obtained their fidelities. 
Finally, when working with Pauli dynamical maps,
we searched for a way of simplifying the parametrized circuit
by requiring that only one single-qubit rotation depends on the parameter.
In theorem \ref{theorem2} we found the general mathematical conditions for 
this, applicable to any parametrized circuit.

Therefore, this work presents yet another example of the current
exploration into simulating open quantum
systems in quantum computers,
and we observe the big effect that the error of quantum computers have
on these simulations.
On the other hand, the result
of theorem \ref{theorem2}
shows what can be done with the condition of using only one
parametrized rotation and
can be applied to any quantum algorithm that requires parametrized circuits,
such as those used for quantum machine learning.

\section{Acknowledgments} 
Support by projects CONACyT 285754, and UNAM-PAPIIT IG101421 is acknowledged.

%
%
\bibliography{Bibliography}

\begin{thebibliography}{10}

\bibitem{feynman1982simulating}
Feynman RP.
\newblock Simulating physics with computers.
\newblock Int J Theor Phys. 1982;21(6/7):467--488.

\bibitem{Zur91}
Zurek WH.
\newblock Decoherence and the transition from quantum to classical.
\newblock Phys Today. 1991;44(10):36--44.

\bibitem{RevModPhys.76.1267}
Schlosshauer M.
\newblock Decoherence, the measurement problem, and interpretations of quantum
  mechanics.
\newblock Rev Mod Phys. 2005;76(4):1267--1305.

\bibitem{breuer2007theory}
Breuer HP, Petruccione F.
\newblock The theory of open quantum systems.
\newblock 1st ed. New York: Oxford University Press; 2002.

\bibitem{Garcia}
García-Pérez G, Rossi M, Maniscalco S.
\newblock IBM Q Experience as a versatile experimental testbed for simulating
  open quantum systems.
\newblock npj Quantum Inf. 2020;6(1):1--11.

\bibitem{Wang}
Wang H, Ashhab S, Nori F.
\newblock Quantum algorithm for simulating the dynamics of an open quantum
  system.
\newblock Phys Rev A. 2011;83(6):062317.

\bibitem{Weimer}
Weimer H, Kshetrimayum A, Or\'us R.
\newblock Simulation methods for open quantum many-body systems.
\newblock Rev Mod Phys. 2021;93(1):015008.

\bibitem{Xin}
Xin T, Wei SJ, Pedernales JS, Solano E, Long GL.
\newblock Quantum simulation of quantum channels in nuclear magnetic resonance.
\newblock Phys Rev A. 2017;96(6):062303.

\bibitem{Wei}
Wei S, Xin T, Long G.
\newblock Efficient universal quantum channel simulation in IBM's cloud quantum
  computer.
\newblock Sci China Phys Mech Astron. 2018;61(7):70311.

\bibitem{Zanetti}
Zanetti M, Pinto D, Basso M, Maziero J.
\newblock Simulating noisy quantum channels via quantum state preparation
  algorithms.
\newblock Phys B At Mol Opt Phys. 2023;56(11):115501.

\bibitem{Andrea}
Far\'{\i}as OJ, Aguilar GH, Vald\'es-Hern\'andez A, Ribeiro PHS, Davidovich L,
  Walborn SP.
\newblock Observation of the emergence of multipartite entanglement between a
  bipartite system and its environment.
\newblock Phys Rev Lett. 2012;109(15):150403.

\bibitem{Andrea_AD}
Aguilar GH, Vald\'es-Hern\'andez A, Davidovich L, Walborn SP, Souto~Ribeiro PH.
\newblock Experimental entanglement redistribution under decoherence channels.
\newblock Phys Rev Lett. 2014;113(24):240501.

\bibitem{Barreiro}
Barreiro J, Müller M, Schindler P, Nigg D, Monz T, Chwalla M, et~al.
\newblock An open-system quantum simulator with trapped ions.
\newblock Nature. 2011;470(7335):486--491.

\bibitem{Marsden}
Head-Marsden K, Krastanov S, Mazziotti D, Narang P.
\newblock Capturing non-Markovian dynamics on near-term quantum computers.
\newblock Phys Rev Res. 2021;3:013182.

\bibitem{chuangbook}
Nielsen MA, Chuang IL.
\newblock Quantum computation and quantum information: 10th anniversary
  edition.
\newblock 10th ed. New York: Cambridge University Press; 2011.

\bibitem{geometry}
Bengtsson I, {\. Z}yczkowski K.
\newblock Geometry of quantum states: an introduction to quantum entanglement.
\newblock 1st ed. Cambridge: Cambridge University Press; 2006.

\bibitem{Zbigniew}
Puchała Z, Łukasz Rudnicki, Życzkowski K.
\newblock Pauli semigroups and unistochastic quantum channels.
\newblock Phys Lett A. 2019;383(20):2376–2381.

\bibitem{Davalos}
Davalos D, Ziman M, Pineda C.
\newblock Divisibility of qubit channels and dynamical maps.
\newblock Quantum. 2019;3:144.

\bibitem{Flammia}
Flammia S, Wallman J.
\newblock Efficient estimation of Pauli channels.
\newblock ACM Transactions on Quantum Computing. 2020;1(3):1--32.

\bibitem{cerezo}
Cerezo M, Arrasmith A, Babbush R, Benjamin SC, Endo S, Fujii K, et~al.
\newblock Variational quantum algorithms.
\newblock Nat Rev Phys. 2020;3(9):625--644.

\bibitem{Benedetti}
Benedetti M, Lloyd E, Sack S, Fiorentini M.
\newblock Parameterized quantum circuits as machine learning models.
\newblock Quantum Sci Technol. 2019;4(4):043001.

\bibitem{Rasmussen}
Rasmussen S, Loft N, Bækkegaard T, Kues M, Zinner N.
\newblock Reducing the amount of single‐qubit rotations in VQE and related
  algorithms.
\newblock Adv Quantum Technol. 2020;3:2000063.

\bibitem{Rieffel}
Rieffel E, Polak W.
\newblock Quantum computing: a gentle introduction.
\newblock 1st ed. Massachusetts: The MIT Press; 2011.

\bibitem{zimansbook}
Heinosaari T, Ziman M.
\newblock {The Mathematical Language of Quantum Theory: From Uncertainty to
  Entanglement}.
\newblock 2nd ed. Cambridge: Cambridge University Press; 2012.

\bibitem{cirac}
Wolf MM, Cirac JI.
\newblock Dividing quantum channels.
\newblock Comm Math Phys. 2008;279(1):147--168.

\bibitem{choi}
Choi MD.
\newblock Completely positive linear maps on complex matrices.
\newblock Linear Algebra Appl. 1975;10(3):285--290.

\bibitem{jamil}
Jamiołkowski A.
\newblock Linear transformations which preserve trace and positive
  semidefiniteness of operators.
\newblock Rep Math Phys. 1972;3(4):275--278.

\bibitem{Marinescu}
Marinescu D, Marinescu G.
\newblock Classical and Quantum Information.
\newblock 1st ed. Oxford: Elsevier; 2012.

\bibitem{Terhal}
Terhal BM.
\newblock Quantum error correction for quantum memories.
\newblock Rev Mod Phys. 2015;87(2):307--346.

\bibitem{Qiskit}
{Qiskit contributors}. Qiskit: An Open-source Framework for Quantum Computing;
  2023.

\bibitem{Chuang:1996}
Chuang IL, Nielsen MA.
\newblock {Prescription for experimental determination of the dynamics of a
  quantum black box}.
\newblock J Mod Opt. 1997;44:2455--2467.

\bibitem{wildebook}
Wilde M.
\newblock From classical to quantum Shannon theory.
\newblock 2nd ed. Cambridge: Cambridge University Press; 2019.

\bibitem{Watrous}
Watrous J.
\newblock Simpler semidefinite programs for completely bounded norms.
\newblock Theor Comput Sci. 2013;19(8):1--19.

\bibitem{Benenti}
Benenti G, Strini G.
\newblock Computing the distance between quantum channels: usefulness of the
  Fano representation.
\newblock J Phys B At Mol Opt Phys. 2010;43(21):215508.

\bibitem{Andrea_4qb}
Aguilar GH, Far\'{\i}as OJ, Vald\'es-Hern\'andez A, Souto~Ribeiro PH,
  Davidovich L, Walborn SP.
\newblock Flow of quantum correlations from a two-qubit system to its
  environment.
\newblock Phys Rev A. 2014;89(2):022339.

\end{thebibliography}
\appendix
\end{document}